\documentclass[aps,pra,twocolumn,floatfix,superscriptaddress]{revtex4-2}
\usepackage{amsmath,amssymb,graphicx,bbold}
\usepackage{graphicx}  
\usepackage{dcolumn}   
\usepackage{bm}        
\usepackage{color}
\usepackage{todonotes}
\usepackage{xcolor}

\usepackage{xcolor}
\usepackage{physics}
\usepackage{ulem}
\usepackage{xspace}
\newcommand{\eg}{e.g.\xspace}
\usepackage{subfigure}

\providecommand{\abs}[1]{\lvert#1\rvert}

\providecommand{\bra}[1]{\langle #1 \rvert}
\providecommand{\ket}[1]{\lvert #1 \rangle}
\providecommand{\braket}[2]{\langle #1 \rvert #2 \rangle}

\usepackage{comment}
\DeclareMathOperator{\sinc}{sinc}

\begin{document}

	\title{Coherence and Dimensionality Witnesses for Fractional OAM Modes}

\author{A. L. S. Santos Junior}
\address{Instituto de F\'{i}sica, Universidade Federal Fluminense, 24210-346 Niter\'{o}i, RJ, Brazil}

\author{I. Prego}
\address{Instituto de F\'{i}sica, Universidade Federal Fluminense, 24210-346 Niter\'{o}i, RJ, Brazil}

\author{M. Gil de Oliveira}
\address{Instituto de F\'{i}sica, Universidade Federal Fluminense, 24210-346 Niter\'{o}i, RJ, Brazil}

\author{A. C. Barbosa}
\address{Instituto de F\'{i}sica, Universidade Federal Fluminense, 24210-346 Niter\'{o}i, RJ, Brazil}

\author{B. P. da Silva}
\address{Instituto de F\'{i}sica, Universidade Federal Fluminense, 24210-346 Niter\'{o}i, RJ, Brazil}
\address{QuIIN—Quantum Industrial Innovation, EMBRAPII CIMATEC Competence Center in Quantum Technologies,
SENAI CIMATEC, Av. Orlando Gomes, 1845, Salvador, Bahia, 41850-010, Brazil}

\author{D. J. Brod}
\address{Instituto de F\'{i}sica, Universidade Federal Fluminense, 24210-346 Niter\'{o}i, RJ, Brazil}

\author{E. F. Galvão}
\address{Instituto de F\'{i}sica, Universidade Federal Fluminense, 24210-346 Niter\'{o}i, RJ, Brazil}
\address{International Iberian Nanotechnology Laboratory (INL) Av. Mestre José Veiga s/n, 4715-330 Braga, Portugal}

\author{A. Z. Khoury}
\address{Instituto de F\'{i}sica, Universidade Federal Fluminense, 24210-346 Niter\'{o}i, RJ, Brazil}

	\date{\today}
	
\begin{abstract}

We characterize sets of fractional orbital angular momentum (OAM) modes of a light beam using unitary-invariant properties encoded by two-mode overlaps. Using basis-independent coherence and dimension witnesses, we experimentally certify, on a triple of fractional modes, both the presence of coherence and that the states necessarily span a space of dimension 3. We propose and implement a practical, fast experimental method to extract two-mode overlaps requiring only a single intensity image per interference pair. These results lay the groundwork for using fractional OAM states in high-dimensional quantum information protocols.

\end{abstract}
	
	\maketitle
    \label{intro}

\section{Introduction}
\label{sec:introduction}

In the paraxial approximation, the total angular momentum of a light beam splits into spin angular momentum and orbital angular momentum (OAM), which arises from the azimuthal phase dependence of the field. While the spin per photon is limited to \(\pm\hbar\), the OAM can take any integer multiple \(l\hbar\) and is characterized by helical wave-fronts of the form \(\exp(i\,l\phi)\) with \(l\in\mathbb{Z}\). A special class of OAM-carrying beams is the family of Laguerre–Gaussian (LG) modes, produced with spiral phase plates\,\cite{beijersbergen1994helical} or holograms\,\cite{bazhenov1990laser}, which impose a phase jump of \(2\pi l\). Non-integer phase jumps have also been demonstrated using plates that impart a phase \(\exp(i\ell\phi)\) with \(\ell\in\mathbb{R}\), thereby introducing a phase discontinuity direction in the transverse plane \cite{berry2004optical}. Because such modes are not solutions of the paraxial wave equation, their transverse intensity profile is unstable upon propagation. Götte \textit{et al.} \cite{Gotte2007,Gotte2008} proposed the controlled synthesis of fractional modes by making a superposition of Laguerre-Gaussian modes, forming a periodically varying pattern along propagation. This method allows the use of 
fractional OAM modes in applications such as optical tweezers for micro-particle manipulation\,\cite{tao2005fractional}, higher-capacity optical communications\,\cite{liu2019superhigh}, and high-dimensional entanglement generation\,\cite{oemrawsingh2004observe,oemrawsingh2005experimental}.

Coherence is an essential resource in quantum metrology \cite{Giovannetti2011}, quantum computing \cite{nielsen2010quantum}, quantum thermodynamics \cite{Lostaglio2015NC}, and quantum biology \cite{Engel2007}. Owing to this relevance, resource theories have been developed to identify and quantify coherence using coherence witnesses \cite{baumgratz2014quantifying}. Such witnesses are observables whose expectation values indicate superposition, identifying that a density operator is not diagonal in the basis set by the observable \cite{baumgratz2014quantifying,napoli2016robustness}. Recently, Galvão and Brod \cite{GalvaoBrod2020} proposed new inequalities formulated in terms of overlaps between pairs of states within a set of at least 3 states. The authors discuss using these inequalities as basis-independent coherence witnesses and as Hilbert-space dimension witnesses. A particularly interesting aspect of the method is that the resulting criteria are fully robust under unitary transformations, apply also to mixed states, and can certify properties of coherence and state dimensionality in a basis-independent way. 
The criterion depends only on the pairwise overlaps between states, whose measurements are experimentally feasible even in scenarios featuring multiple states.

Understanding the coherence properties and dimensionality of the Laguerre–Gaussian (LG) and Hermite–Gaussian (HG) mode families has enabled the use of high dimensionality as a resource in various applications \cite{Dunlop2016}, such as optical communications \cite{willner2021orbital,nape2023quantum} and quantum information tasks \cite{miao2022high}. For these mode families, certain structural properties of sets of states, particularly the physical Hilbert space they generate (i.e., the dimensionality of the mode set) 
are well established: LG and HG form complete, orthonormal bases for the transverse optical field. By contrast, this does not immediately apply to beams carrying fractional orbital angular momentum, for which completeness, orthogonality, and consequently the dimensionality of the space spanned by a set of modes are not trivially determined. The inner product of two fractional modes already reveals this additional complexity: the overlap is a nontrivial function of $\ell$ and of the relative orientation of the phase discontinuity, introducing an extra parameter. Therefore, investigating the relational properties of fractional modes, revealing their structure, can enable more effective exploitation of their properties in various communication and information processing tasks.

In this scenario, we propose a characterization of fractional modes 
based on the pairwise overlaps observed among them. In this work, we apply coherence and dimension witnesses to characterize previously unexplored relational properties of fractional modes. We consider three fractional modes and state conditions on two-state overlaps that guarantee the three states cannot be simultaneously diagonal, thus ensuring they display coherence.  We also identify conditions on overlaps which guarantee the states must span at least a three-dimensional  space. To validate our analysis, we implement an experimental method that retrieves the overlaps between spatial modes from a single image of the intensity distribution resulting from the interference of each pair of states. The measured values show good agreement with theoretical predictions. These results are particularly relevant for studies of photonic indistinguishability, for high-dimensional quantum key distribution (QKD) and other quantum communication protocols.

This paper is structured as follows. Section~\ref{sec:Coherence_and_Dimensionality} reviews coherence and dimensionality witnesses for sets of modes. Section~\ref{sec:Mode_overlap_measurements} proposes an interferometric method to measure pairwise overlaps between spatial modes via transverse superposition. We also experimentally validate the reliability of our apparatus by measuring overlaps between pairs of fractional modes, corroborating theoretical expectations. In Section~\ref{sec:Dimension_and_Coherence_of_Fractional_OAM_Modes}, we first test these witnesses on a set of three integer-valued OAM modes. Next, we apply them to a different configuration of three fractional modes with topological charge restricted to $0\leq \ell \leq 1$ and experimentally demonstrate the conditions under which they display basis-independent coherence and high dimensionality. Finally, Section~\ref{sec:conclusion} presents our conclusions.

\section{Review: Coherence and Dimensionality Witnesses}
    \label{sec:Coherence_and_Dimensionality}
    
Coherence is an essential resource behind paradigmatic physical phenomena, such as interference. In quantum mechanics, coherence is often defined with respect to a reference basis, and determining whether a state displays it is a nontrivial task, requiring specific observables to witness the fact the state cannot be diagonalized in that basis \cite{Quantitative_coherence_witness2017,napoli2016robustness}. In  \cite{GalvaoBrod2020} the authors proposed an alternative approach, capable of using only overlaps between pairs of states in a set to witness a notion of basis-independent coherence. More concretely, they proposed a set of inequalities for two-state overlaps, when violated, implies that there is no basis in Hilbert space that simultaneously diagonalizes all states in the set. We now give a brief description of the theoretical proposal of this coherence witness, which we use to experimentally characterize fractional OAM states in subsequent sections. 

Consider some arbitrary quantum state $\rho$ and some observable $\hat{O}$, with basis of eigenstates $\{\ket{\phi_i}\}_{i=1 \ldots n}$. If $\rho$ is diagonal in that basis it defines a classical probability distribution, $\rho = \sum_i p_i\,|\phi_i\rangle\langle\phi_i|$, and we say that $\rho$ is incoherent with respect to $\hat{O}$. Measuring $\hat{O}$ then directly samples from the probability distribution $\{p_i\}$. 

Let $\{\rho_i\}_{i=1 \ldots n}$ be a set of arbitrary quantum states. Suppose two of these states, say $\rho_i$ and $\rho_j$, are incoherent with respect to the same observable $\hat{O}$. Then the overlap between these two states, given by $r_{ij} = \mathrm{tr}(\rho_i \rho_j)$, can be interpreted as the (classical) probability that both states yield an identical outcome in independent measurements of $\hat{O}$.

By treating these incoherent (with respect to $\hat{O}$) states as classical probability distributions, we can derive bounds, as done in \cite{GalvaoBrod2020}, that must be satisfied by sets of pairwise overlaps. For example, if we have three states $\{\rho_A, \rho_B, \rho_C\}$, we can show that

\begin{equation}\label{eq:classical_bound}
\begin{aligned}
  r_{AB} + r_{BC} - r_{AC} &\le 1,\\
  r_{AB} - r_{BC} + r_{AC} &\le 1,\\
  -r_{AB} + r_{BC} + r_{AC} &\le 1.
\end{aligned}
\end{equation}

The crucial observation in \cite{GalvaoBrod2020} is that whenever a set of three-state overlaps violates one of these inequalities, no basis exists in which the states are simultaneously incoherent. Furthermore, overlaps are invariant under changes of basis, which means that, as long as we have experimental access to these quantities, this fact can be tested even if we have zero knowledge of which basis diagonalizes any of the states. Therefore, these inequalities act as basis-independent witnesses of coherence.

The fact that some quantum states violate the bounds of \eqref{eq:classical_bound} can be illustrated by the following triple of states,  

\begin{equation}
    \left\{\ket{0}, \frac{1}{2}\left(\ket{0}+\sqrt{3} \ket{1}\right), \frac{1}{2}\left(\sqrt{3} \ket{0}+\ket{1}\right)\right\}. 
\end{equation}

However, not every triple of overlaps is physically realizable in quantum mechanics. For instance, $(r_{AB}, r_{BC}, r_{AC}) = (1,1,0)$ is impossible: $r_{AB}=1$ and $r_{BC}=1$ force $\rho_A$ and $\rho_C$ to coincide with $\rho_B$ (up to a phase), which rules out $r_{AC}=0$. Thus, we expect the region of overlaps compatible with quantum mechanics to also be described by some bounds, and \cite{GalvaoBrod2020} indeed have shown those to be

\begin{equation}
\label{eq:quant_ineq}
r_+ \geq r_{BC} \geq 
\begin{cases}
r_-, & \text{if } r_{AB} + r_{AC} > 1, \\
0,   & \text{otherwise,}
\end{cases}
\end{equation}

where

\begin{equation}
r_\pm := \left(\sqrt{r_{AB}\,r_{AC}} \,\pm\, \sqrt{(1 - r_{AB})(1 - r_{AC})}\right)^2,
\end{equation}

and with analogous bounds obtained from relabeling of the three states.

Finally, we can obtain further constraints on the overlaps if there is a restriction on the dimensionality of the underlying state space. For instance, the situation where all three states are orthogonal (i.e., all overlaps are 0) is possible when the states live in a Hilbert space of dimension greater than or equal to 3, but not for qubits. Therefore, triples of overlaps can also serve as dimension witnesses, if we measure them to be in a region incompatible with a lower-dimensional system.

\begin{figure}
    \centering
    \includegraphics[width=1.0\linewidth]{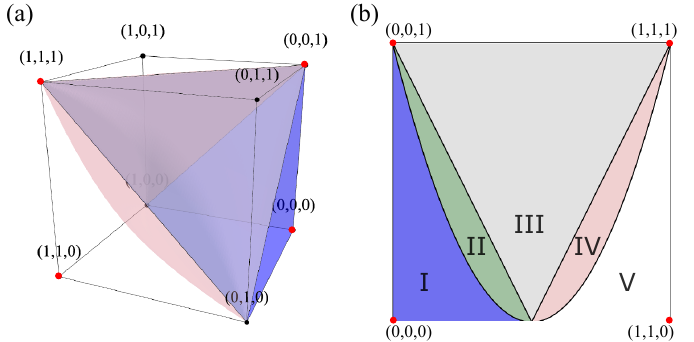}
    \caption{Classical and quantum bounds for the three pairwise overlaps among three states, parameterized by $(r_{AB}, r_{BC}, r_{AC})$. In (a), the blue polyhedron represents \(\mathcal{C}\), while the curved pink surface corresponds to the boundaries of the quantum region \(\mathcal{Q}\). In (b), we display the cross section in the case $r_{AB} = r_{BC}$. Regions I, II, and III belong to \(\mathcal{C}\), while region IV belongs exclusively to \(\mathcal{Q}\).}
    \label{fig:fig_bounds}
\end{figure}

We collect all constraints derived in \cite{GalvaoBrod2020} in Figure \ref{fig:fig_bounds}. We denote by $\mathcal{C}$ the solid region consistent with classical probability distributions (i.e., all states incoherent in a common basis), defined by \eqref{eq:classical_bound}, and by $\mathcal{Q}$ the region consistent with arbitrary quantum states, i.e., region defined by \eqref{eq:quant_ineq}. We also denote by $\mathcal{Q}_{bid}$ the region compatible with the overlaps of three qubits. Figure \ref{fig:fig_bounds}(a) represents the full three-dimensional space accessible to overlap vectors $\mathbf{r} = (r_{AB}, r_{BC}, r_{AC})$.

To illustrate all the possibilities identified so far, Fig\ \ref{fig:fig_bounds}(b) represents a two-dimensional slice where \(r_{AB}=r_{BC}\). In region I, the overlaps act as dimension witnesses, as they are only compatible with states that span a subspace of dimension \(d > 2\). Region II is accessible to either two-dimensional quantum states or three-dimensional classical (i.e., coherence-free) states. Region III is compatible with general classical probability distributions, even on two-dimensional configuration spaces. Region IV serves as a witness of coherence, because it lies in \(\mathcal{Q}\) but not in \(\mathcal{C}\). Finally, region V cannot be reached by any set of three states, as it lies outside \(\mathcal{Q}\).

Subsequent work showed further coherence and dimensionality witnesses for sets of \(n\) states of dimension \(d\) \cite{giordani2023experimentalcertification}. The witnesses are defined by violations of a family of inequalities defined recursively in \cite{giordani2023experimentalcertification} by:
\begin{equation}\label{hn_coherence_witness}
   h_n(r) = h_{n-1}(r) + r_{0,n-1} - \sum_{i=1}^{n-2} r_{i,n-i} \;\le\; 1\,,
\end{equation}
with
$
h_3(r) = r_{01} + r_{02} - r_{12},
$
which coincides with one of the inequalities in Eq.~\eqref{eq:classical_bound}. The functions \(h_n\) are specifically designed to witness coherence in systems of dimension \(d>2\) through pairwise overlaps. Violation of the inequality \(h_n(r)\leq 1\) constitutes an unequivocal witness of quantum coherence, as \eg{no basis can simultaneously diagonalize all \(n\) states}. Moreover, it is shown that no set of incoherent states, nor any system of dimension \(d<n-1\), exceeds the bound \(h_n=1\); that is, two-dimensional systems do not violate \(h_4\),  three-dimensional systems do not violate \(h_5\), and so on. The maximum achievable value of \(h_n\) depends exclusively on the dimension of the Hilbert space, so that the functions \(h_n(r)\) themselves act simultaneously as witnesses of coherence and of the minimum dimensionality required to reproduce that degree of violation.

	\section{Paraxial mode overlap measurements}
	\label{sec:Mode_overlap_measurements}

 In this section, we introduce a method for measuring overlaps between transverse modes of paraxial beams, closely related to the approaches used in Refs. \cite{nishchal2006binary,shikder2024fractional}. We begin by defining fractional OAM and deriving the overlap between two arbitrary fractional-OAM states. We then present our overlap-measurement scheme and validate it on fractional modes, comparing the measured overlaps with the theoretical predictions obtained.

\subsection{Fractional OAM}

We begin by reviewing some properties of fractional orbital angular momentum states. Recall that Laguerre-Gaussian (LG) transverse modes are solutions of the paraxial wave equation in free space and correspond to the eigenmodes of optical resonators with cylindrical symmetry. These quantized modes carry an OAM of \(l\hbar\) per photon, with \(l \in \mathbb{Z}\). In the position representation, they can be expressed as
\begin{align}
\braket{r,\phi}{l,p} &= u_{lp}(r,\phi)  \notag \\
& = \frac{\mathcal{N}_{lp}}{\sqrt{2\pi}}\, \bar{r}^{|l|}\, 
L^{|l|}_p(2\bar{r}^2)\, e^{-\bar{r}^2}\, e^{i l\phi},\label{LG_modes}
\end{align}
where \(\mathcal{N}_{lp}\) is a normalization constant, \(L^{|l|}_p\) is the generalized Laguerre polynomial, and \(\bar{r} = r/w\) is the normalized radial coordinate. These modes form a complete orthonormal basis, with \(\left|\braket{l,p}{l',p'}\right|^2 = \delta_{l,l'}\,\delta_{p,p'}\).

Fractional OAM modes were first created by introducing a phase plate with a phase step in the path of a regular LG beam \cite{oemrawsingh2004observe,oemrawsingh2005experimental}. The role played by this phase plate is essentially to introduce a linear azimuthal phase shift \(\ell(\phi - \alpha)\), with charge parameter \(\ell \in \mathbb{R}\) and a phase step at angle \(\alpha\). The parameter can be written in terms of its integer and fractional parts, \(\ell = j + \lambda\), where \(j \in \mathbb{Z}\) and \(\lambda \in\, ]0,1[\).

The action of the phase plate with a step can be described by the unitary operator \(S(\ell,\alpha)\), which in the position representation is given by

\begin{align*}
    \bra{r,\phi} S(\ell,\alpha)\ket{l,p} &=  u_{lp}(r,\phi)\,f_{\ell,\alpha}(\phi), \notag \\
    f_{\ell,\alpha}(\phi) &= e^{i\ell(\phi-\alpha)} \times
    \begin{cases}
    e^{i\,2\pi\ell}, & 0 \le \phi < \alpha\\
    1, & \alpha \le \phi < 2\pi
    \end{cases}
    \,.
\end{align*}

We can then formally define the fractional OAM states as

\begin{eqnarray}
    \ket{l,p;\ell,\alpha} &=& S(\ell,\alpha)\ket{l,p}\,.
    \label{eq:fracOAMstate}
\end{eqnarray}

To use the witnesses introduced in Section \ref{sec:Coherence_and_Dimensionality} we need the overlaps for sets of fractional OAM modes, \(\braket{l',p';\ell',\alpha'}{ l,p;\ell,\alpha}\). We restrict our analysis to the cases \(l = l'\) and \(p = p'\), which are sufficient to capture the essential properties of these states. Since our results do not depend on \(l\) and \(p\), we omit these indices from now on and simply write

\begin{equation}
\braket{\ell',\alpha'}{\ell,\alpha} 
= \frac{1}{2\pi} \int_0^{2\pi} f^*_{\ell',\alpha'}(\phi)\, f_{\ell,\alpha}(\phi)\,d\phi \,.
\label{eq:fracOverlap2}
\end{equation}

From the derivation presented in Appendix~A, the absolute value of the overlap is readily obtained as

\begin{align}
\label{overlap_frac}
\abs{\braket{\ell',\alpha'}{\ell,\alpha}}^2 = & 
        		\sinc^2[(\ell - \ell')\pi] - \frac{\beta(2\pi-\beta)}{\pi^2} S_{\ell \ell'}(\beta),  \notag
        				\\
S_{\ell \ell'}(\beta) =& \sin(\ell\pi)\sin(\ell'\pi)\,
        		\sinc {\left[(\ell - \ell')\frac{\beta}{2}\right]}
            \notag \\
        	&\times	\sinc\left[(\ell - \ell')\left(\pi-\frac{\beta}{2}\right)\right]\,,
        	\end{align}
where \(\beta = \alpha - \alpha'\). Note that \(\abs{\braket{\ell',\alpha'}{\ell,\alpha}}^2 = \abs{\braket{\ell',0}{\ell,\beta}}^2\).
    
\subsection{Overlap Measurements}

\begin{figure}[b]
    \centering
    \includegraphics[width=1\linewidth]{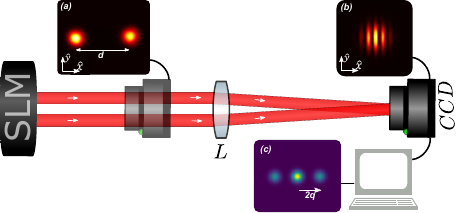}
    \caption{We present our experimental setup. First, we generate two independent modes, separated by a distance $d$ along $\hat{x}$, using a spatial light modulator (SLM). Figure~(a) shows, via a CCD camera, an image of two identical fundamental Gaussian modes. A lens (L) is positioned to introduce an additional transverse momentum component along $\hat{x}$, producing in the focal plane the interference pattern shown in Figure~(b). We then compute the Fourier transform of the pattern in (b); the result is shown in Figure~(c), where the relevant information is concentrated at the spatial frequency $(\pm 2q,0)$. } 
    \label{fig:setup}
\end{figure}

We use a Nd:YAG laser with $1064$ nm wavelength to illuminate a spatial light modulator (SLM), which generates two separate transverse modes $u_A(\mathbf{r})$ and $u_B(\mathbf{r})$ spaced by $d = 0.5$ cm in the \(\hat{x}\)-direction. The experimental setup is presented in Figure ~\ref{fig:setup}.

A lens \(L\) with a focal length of $30$ cm is positioned such that both modes are incident upon points equidistant from its center. This configuration introduces a linear phase factor, \(\exp(\pm iqx)\), causing the beam to acquire a transverse momentum component \(\pm q\) whose magnitude depends on \(d/2\) and the focal length of lens \(L\). Thus, in the lens’s focal plane, the beams undergo a lateral shift, causing their intensity distributions to overlap and interfere with one another. The intensity distribution in the interference plane is given by $I = I_A + I_B + I_{AB} + I^{*}_{AB}\,,$
where $I_{j}(\mathbf{r}) = |u_j(\mathbf{r})|^2$ ($j=A,B$) and

\begin{align}
I_{AB}(\mathbf{r}) = & u_A(\mathbf{r}) u_B^*(\mathbf{r}) \exp(-2iqx)\,.
\end{align}

Let $\tilde{I}(\mathbf{k}) = \mathcal{F}[I(\mathbf{r})]$ be the two-dimensional Fourier transform 
of the interference pattern. Note that $r_{AB} = |\tilde{I}_{AB}(2q,0)|^2\,$.
When the transverse momentum kick $q$ is sufficiently larger than the 
width of $\mathcal{F}[I_{j}(\mathbf{r})]\,$, the supports of $\mathcal{F}[I_{AB}(\mathbf{r})]\,$, 
$\mathcal{F}[I_{AB}^*(\mathbf{r})]$ and $\mathcal{F}[I_{j}(\mathbf{r})]$ lie in disjoint domains 
of the Fourier space. In this case, the overlap is simply given by
\begin{equation}
\label{overlap_comp}
r_{AB} = \left|\int u_A^*(\mathbf{r})\,u_B(\mathbf{r})\,d^2\mathbf{r}\right|^2 
\simeq \left|\tilde{I}(\pm 2q,0)\right|^2\,.
\end{equation}
The resolution of $I_j\,$, $I_{AB}$ and $I^*_{AB}$ in Fourier domain could be checked experimentally by observing the Fourier space image in real time, as we show next with a Gaussian test beam.

To validate this experimental method for measuring overlaps, we first produced two Gaussian modes with a beam waist of $w=1\,\mathrm{mm}$, as illustrated in Fig.~\ref{fig:setup}(a), and recorded them with a CCD camera. Figure~\ref{fig:setup}(b) shows the experimentally obtained interference pattern between these modes, while Fig.~\ref{fig:setup}(c) displays the Fourier transform calculated numerically with Python. The identification in Eq.~\eqref{overlap_comp} is valid only if $q$ is chosen so that the contributions from the distributions centered at $0$ and $-2q$ are negligible (or zero) at the frequency $(2q,0)$. This condition is satisfied in our experiment, as evidenced by Fig.~\ref{fig:setup}(c), where the function supports are sufficiently separated. Consequently, we obtained an overlap of $0.98 \pm 0.01$ after 100 repetitions. The inability to achieve the theoretical overlap of~1 is due to imperfections in the mode preparation by the SLM, which prevent the production of two perfectly identical modes.

\begin{figure*}
    \centering
    \includegraphics[width=0.8\linewidth]{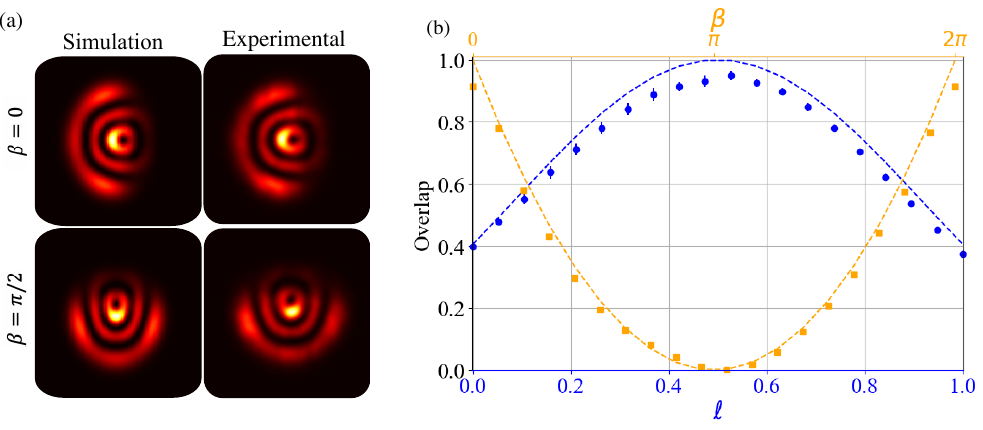}
    \caption{(a)  Fractional modes obtained by superposing \(n_{\mathrm{modes}}=10\) for \(\beta=0\) and \(\beta=\frac{\pi}{2}\) are compared with simulations. (b) Performance evaluation of the overlap method: in blue, the modes have a fixed phase jump (\(\beta=0\)) with \(\ell'=1/2\) and \(\ell\) varying from 0 to 1; in orange, both modes have fixed fractional charge (\(\ell=\ell'=1/2\)) and the angle \(\beta\) varies from 0 to \(2\pi\).}
    \label{fig:fractional_modes.pdf}
\end{figure*}

Using the approach described by \cite{gotte2008light}, we generated modes with fractional charge, \(\ell = 1/2,\) by superposing \(n_{\mathrm{modes}} = 10\) modes. The experimental results of the intensity distribution are presented in Figure~\ref{fig:fractional_modes.pdf}(a) for \(\beta = 0\) and \(\beta = \pi/2\). In Figure~\ref{fig:fractional_modes.pdf}(b), we experimentally evaluate the overlap between two fractional modes in two distinct regimes. In the first situation, we fix \(\beta=0\), \(\ell'=1/2\) and vary \(\ell\) from 0 to 1. The theoretical overlap given by \eqref{overlap_frac} then reduces to \(\sinc^2[\pi(\ell-1/2)]\), represented by the dashed blue curve. The blue dots correspond to the respective experimental results. In the second regime, with \(\ell=\ell'=1/2\), the overlap between the modes is given by \(\bigl(1-\beta/\pi\bigr)^2\), shown by the dashed orange line in the same Figure~\ref{fig:fractional_modes.pdf}(b). The experimental points were obtained by varying \(\beta\) from 0 to \(2\pi\), indicated by the orange squares. Both show good agreement with the theoretical curves.

\section{Dimension and Coherence of Fractional OAM Modes}
\label{sec:Dimension_and_Coherence_of_Fractional_OAM_Modes}

In this Section we discuss how to apply the inequalities described in Section \ref{sec:Coherence_and_Dimensionality} to infer relational properties of fractional modes. These modes display a rich phase structure that makes it difficult to anticipate the behavior of the achievable overlaps and corresponding witnesses.

\begin{figure}
    \centering
    \includegraphics[width=0.9\linewidth]{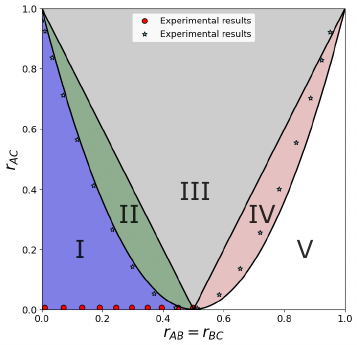}
    \caption{{Two sets of states with integer angular momentum. The red markers indicate the violation of the dimension witness for the triplet, corresponding to a set of three nearly-orthogonal states close to the origin. The blue markers show the violation of the coherence witness. Error bars are small
and therefore not shown in the figure}}
    \label{fig:LGmodes}
\end{figure}

As a first step, we experimentally validated our measurement method and the witnesses described in Sec. \ref{sec:Coherence_and_Dimensionality} using the well-understood integer-OAM modes. To that end, we prepared sets of trios of states

\begin{align*}
\ket{\psi_{A}} & = \cos\theta\,u_{10}
                                  + \sin\theta\,u_{-10}\,, \\
\ket{\psi_{B}} &= u_{10}\,,
\\
\ket{\psi_{C}} &= \cos\theta\,u_{10}
                                  - \sin\theta\,u_{-10}\,, \\
 \theta&\in[0,\pi/2]\,,
\end{align*}

with variable $\theta$,  and which we represent in Fig.\ \ref{fig:LGmodes} by blue stars. For all points falling in Region IV, the triple of states necessarily forms a coherent superposition, which means there is no single basis in which all three states are simultaneously diagonal. Likewise, we generated the states

\begin{align*}
\ket{\psi_{A}} &= u_{20}\,, \\
\ket{\psi_{B}} &= \sqrt{\tfrac{1-\epsilon}{2}}\,
                   (u_{20}+u_{-20})
                 + \sqrt{\epsilon}\,u_{01}\,,\\
\ket{\psi_{C}} & = u_{-20}\,, \\
 \epsilon&\in[0,1],
\end{align*}

which correspond to red dots in Fig.\ \ref{fig:LGmodes}. These are found within region I, which clearly shows that they span a three-dimensional vector space, as predicted by theory.

 Now we apply the dimension and coherence witnesses developed in Section \ref{sec:Coherence_and_Dimensionality} to a set of fractional OAM modes. Consider three fractional OAM states,
\begin{eqnarray}
	\ket{\psi_i} = \ket{\ell_i,\alpha_i},
\end{eqnarray}
where $i \in\{A,B,C\}$ and $\alpha_A > \alpha_B > \alpha_C$. The overlaps \(\abs{\braket{\ell_i,0}{\ell_j,\beta}}^2\) are given by Eq.~\eqref{overlap_frac}, with \(\beta_{ij} = \alpha_i - \alpha_j\). Note that \(\beta_{AB} + \beta_{BC} = \beta_{AC}\). The different parameters that characterize the fractional OAM modes allow for a wide variety of overlap conditions. To capture the essential features of these fractional modes, we analyze two illustrative cases in what follows.

\subsection{Experimental witnessing of coherence}
\label{subsec:condition-2}

We analyzed the coherence witness triplets of fractional modes over the full parameter space defined by the fractional charges
\(\ell\) and the phase‐discontinuity orientations \(\beta\), which fix the pairwise overlaps through Eq.~\eqref{overlap_frac}. The largest violation of inequality \eqref{eq:classical_bound} was obtained when all three discontinuities point in the same direction, that is, when \(\beta_{ij}=0\) for every pair of modes. A comprehensive numerical sweep confirms that this symmetric configuration therefore represents the most favorable scenario for probing coherence in fractional modes.

In this situation, the overlaps depend only on the fractional topological charges, and their expressions reduce to:
\begin{equation}
	r_{ij} = \mathrm{sinc}^2[(\ell_i - \ell_j)\pi]\;.
	\label{eq:rij-beta=0}
\end{equation}
Eq.~\eqref{eq:rij-beta=0} shows that the violation condition is invariant under permutations of $\ell_A$, $\ell_B$, and $\ell_C$. Therefore, without loss of generality, we take $\ell_A>\ell_B>\ell_C$. Since the overlaps depend only on the differences $\ell_i-\ell_j$, we can parameterize them by two variables, $x\equiv(\ell_A-\ell_B)\pi$ and $y\equiv(\ell_B-\ell_C)\pi$, so that:

\begin{eqnarray}
	F_1(x,y) = -\mathrm{sinc}^{2}(x) + \mathrm{sinc}^{2}(x+y) + \mathrm{sinc}^{2}(y) \;\;\leq\;\; 1 \,,
	\nonumber\\
	F_2(x,y) = +\mathrm{sinc}^{2}(x) - \mathrm{sinc}^{2}(x+y) + \mathrm{sinc}^{2}(y) \;\;\leq\;\; 1 \,,
	\nonumber\\
	F_3(x,y) = +\mathrm{sinc}^{2}(x) + \mathrm{sinc}^{2}(x+y) - \mathrm{sinc}^{2}(y) \;\;\leq\;\; 1 \,,
	\nonumber\\
	\label{eq:inequalities-xy}
\end{eqnarray}
We can look for parameter values that lead to a violation of these conditions by searching for solutions of \(\boldsymbol{\nabla} F_j = \mathbf{0}\), corresponding to maxima of \(F_j(x,y)\). It can be shown that the only solution to \(\boldsymbol{\nabla} F_1 = \mathbf{0}\) and \(\boldsymbol{\nabla} F_3 = \mathbf{0}\) is \(x = y = 0\). However, there are nontrivial solutions for \(\boldsymbol{\nabla} F_2 = \mathbf{0}\), given by \(y = x\) and solutions of the transcendental equation
\begin{equation}
	4x\sin(2x) - 8\sin^2(x) - x\sin(4x) + \sin^2(2x) = 0\,.
	\label{eq:transcendental}
\end{equation}
The maximum violation obtained was \(F_2(x,y) \approx 1.22\) when \(x = y = 0.28\pi\).

\begin{figure}
    \centering
    \includegraphics[width=1\linewidth]{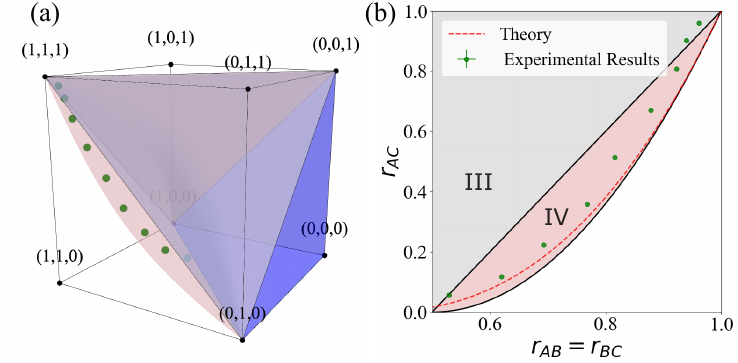}
    \caption{(a) Experimental results for a set of fractional modes described in \eqref{caseone}, with \(\epsilon\) varying from \(0\) to \(1/2\). (b) We project the experimental points onto the plane \(r_{AB} = r_{BC}\), showing that these beams lie  in region~\(\mathcal{Q}\), outside region \(\mathcal{C}\). Error bars are small
and therefore not shown in the figure.}
    \label{fig:exp_coherence_witness}
\end{figure}

Following the condition $x = y$ established above, we choose a set of fractional states defined by

\begin{equation}\label{caseone}
    \{\ket{\psi_A} = \ket{\tfrac{1}{2}-\epsilon,0},  \ket{\psi_B} = \ket{\tfrac{1}{2},0},  \ket{\psi_C} = \ket{\tfrac{1}{2}+\epsilon,0}\},
\end{equation}

with \(\epsilon \in [0,1/2]\). In Fig.\ \ref{fig:exp_coherence_witness}(a), we show the corresponding experimental points within the geometric representation of the inequalities, considering \(\mathbf{r} = (r_{AB}, r_{BC}, r_{AC})\). Clearly, these points lie outside the polyhedron \(\mathcal{C}\) but inside the volume \(\mathcal{Q}\), whose boundaries are defined in Eq.~\eqref{eq:quant_ineq}. For clarity, we project the points and the two boundaries of \(\mathcal{C}\) and \(\mathcal{Q}\) onto the plane \(r_{AB} = r_{BC}\), as illustrated in Fig.~\ref{fig:exp_coherence_witness}(b).
 We see that the points lie in region~IV, demonstrating that these fractional modes exhibit basis-independent coherence.

For the point of maximum predicted violation, we set \(\epsilon=0.28\). We define the degree of violation of inequalities~\eqref{eq:classical_bound}, denoted by \(\boldsymbol{W_c}\), as the Euclidean distance of the triple of overlaps \(\bigl(r_{AB},r_{BC},r_{AC}\bigr)\) to the nearest face of \(\mathcal{C}\). The experimental value, reported in Table~\ref{tab:overlap_s1}, violates the inequality by six standard deviations.

Importantly, the error bars are purely statistical, computed from \(N=100\) repeated measurements. They do not account for systematic imperfections in the mode quality. For instance, the modes must be positioned slightly off-center within the lens aperture, which unavoidably introduces small asymmetries between modes. A further limitation comes from the finite pixel resolution of the SLM, which constrains the fidelity of the synthesized phase profiles. These factors explain the small deviations from the theoretical (red) curve in Fig.~\ref{fig:exp_coherence_witness}. Nevertheless, the observed violations remain significant, by many standard deviations, and the method is robust, requiring no phase locking between mode pairs.

\begin{table}[h]
    \centering
    \caption{Experimental overlap values for the fractional modes \(\ell_A = 0.22\), \(\ell_B = 0.5\), and \(\ell_C = 0.78\) with \(\beta_{AB} = \beta_{BC} = 0\).}
    \label{tab:overlap_s1}
    \begin{tabular}{c c c c}
        \hline
        \(\boldsymbol{r_{AB}}\) & \(\boldsymbol{r_{BC}}\) & \(\boldsymbol{r_{AC}}\) & \(\mathbf{W_c}\) \\
        \hline
         \(0.62 \pm 0.01\) & \(0.71 \pm 0.01\) & \(0.23 \pm 0.01\) & \(0.06 \pm 0.01\) \\
        \hline
    \end{tabular}
\end{table}

\subsection{{Experimental witnessing of dimension}}
\label{subsec:condition-1}

After a numerical search, we identified a family of states which enters region I (i.e., whose pairwise overlaps cannot be explained by a two-dimensional system) with high visibility. Those correspond to the case where all fractional orbital angular momentum modes share the same topological charge, \(\ell_{A}=\ell_{B}=\ell_{C}\equiv\ell\), and the relative
discontinuity angles \(\beta_{ij}\) are free parameters.
Under this condition the pairwise overlaps simplify to  

\begin{equation}
   r_{ij}=1-\frac{\beta_{ij}\left(2\pi-\beta_{ij}\right)}{\pi^{2}}
           \,\sin^{2}\!\left(\ell\pi\right).
   \label{eq:overlap_sym}
\end{equation}

\begin{figure}[h]
    \centering
    \includegraphics[width=1\linewidth]{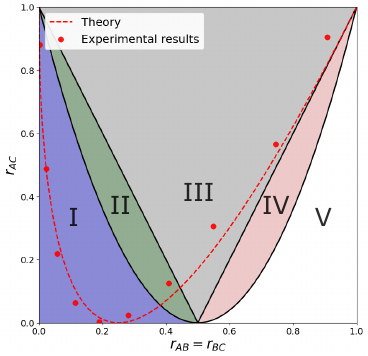}
    \caption{A set of measurements was carried out for the fractional modes in \eqref{modesbetas}, where \(\beta\) ranged from \(0\) to \(2\pi\). Error bars are small
and therefore not shown in the figure}
    \label{fig:regionone}
\end{figure}

To test the prediction that these states cannot be embedded in any two-dimensional subspace, we experimentally prepared the triplet of states

\begin{align}\label{modesbetas}
\ket{\psi_A} &= \ket{\tfrac12,+\beta}, \\
     \ket{\psi_B} &= \ket{\tfrac12,0}, \\
     \ket{\psi_C} &= \ket{\tfrac12,-\beta},
\end{align}

where \(\beta\) varies from \(0\) up to \(\pi\), noting that \(r_{AB} = r_{BC}\). This choice also allows for a two-dimensional visualization, as illustrated in Figure \ref{fig:regionone}. The experimental values are projected onto plane $r_{AB}=r_{BC}$ and shown as red points. A set of overlap values appears in Region I, outside \(\mathcal{Q}_{\mathrm{bid}}\), confirming that the states must span a three-dimensional Hilbert space. We define the dimension witness \(\mathbf{W}_D\) as the maximum Euclidean distance between the measured overlap values and the region \(\mathcal{Q}_{\mathrm{bid}}\). Our result,
shown in Table \ref{tab:overlap_s2}, is \(\mathbf{W}_D = 0.31 \pm 0.01,\) which violates \(\mathcal{Q}_{\mathrm{bid}}\) by about 30 standard deviations.

\begin{table}
    \centering
    \caption{The measured triple of overlap values maximally violates the volume defined by \(\mathcal{Q}\), which necessarily span a three-dimensional space.}
    \label{tab:overlap_s2}
    \begin{tabular}{c c c c}
        \hline
        \(\boldsymbol{r_{AB}}\) & \(\boldsymbol{r_{BC}}\) & \(\boldsymbol{r_{AC}}\) & \(\mathbf{W_c}\) \\
        \hline
         \(0.05 \pm 0.01\) & \(0.06 \pm 0.01\) & \(0.11 \pm 0.01\) & \(0.31 \pm 0.01\) \\
        \hline
    \end{tabular}
\end{table}

\subsection{Accessible region for fractional OAM modes}
\label{sec:fracOAMdistinguishability}

The fact that fractional modes are superpositions of infinitely many Laguerre modes is suggestive that they live in a high-dimensional Hilbert space. However, that could be an artifact of the choice of basis they are written in---the space spanned by three modes is always at most three-dimensional, and it is conceivable that they would be linearly-dependent and span a smaller subspace. 

Our results from Sec.\ \ref{subsec:condition-1} show that this is not generally the case, as we can witness linear independence even when the fractional charge is restricted to values between \(0\) and \(1\). However, that does raise an additional theoretical question: can a set of three fractional OAM modes span the entire space accessible to quantum mechanics in Fig.\ \ref{fig:fig_bounds}? 

The answer to that question is trivial if we allow $\beta_{ij}$ and $\ell$ to vary arbitrarily, since in particular this would include OAM states with three distinct \emph{integer} values of $\ell$. Since those consist of three mutually orthogonal states, they do form a full three-dimensional Hilbert space, and thus can naturally span the entire polytope \(\mathcal{Q}\). 

We found a more interesting case when we restrict all three states to have $0 \leq \ell \leq 1$. On one hand, these do not automatically include a set of three orthogonal states, so we might expect them to not span the entire region \(\mathcal{Q}\). On the other, we have already shown that even in this case we can find sets of three states that are linearly independent. 

\begin{figure}[h]
    \centering
    \includegraphics[width=1\linewidth]{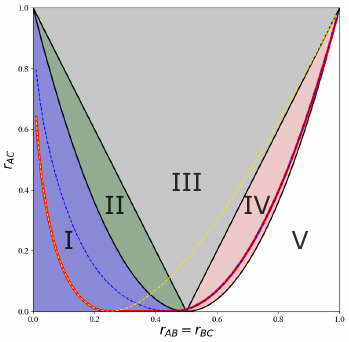}
    \caption{Results of a numerical search over the five free parameters determined for three arbitrary fractional modes.  The solid red curve marks the boundary of overlaps achievable with arbitrary fractional OAM modes where \(\{\ell_1, \ell_2 , \ell_3\} \in [0,1]\). The dashed yellow boundary corresponds to the symmetric case \(\ell_1 = \ell_2 = \ell_3\), as shown in subsection \ref{subsec:condition-1}. The dashed blue boundary corresponds to the family defined by the constraint \(\beta_{AB} + \beta_{AC} = \pi\), with \(\ell_1\), \(\ell_2\), and \(\ell_3\) completely free.}

    \label{fig:new_boundary}
\end{figure}

Indeed, a full numerical search on the set of five parameters $\{\ell_1, \ell_2, \ell_3, \beta_{12}, \beta_{13}\}$, with $0\leq\ell_i\leq 1$ reveals that these states do delineate a novel boundary, distinct from those of \(\mathcal{C}\) and \(\mathcal{Q}\), which we represent in Fig.~\ref{fig:new_boundary} with a solid red line. In this figure we also display in dashed yellow the boundary corresponding to the symmetric case \(\ell_1 = \ell_2 = \ell_3\), as shown in subsection \ref{subsec:condition-1}, and in dashed blue the boundary corresponding to the set defined by the constraint \(\beta_{AB} + \beta_{AC} = \pi\), with \(\ell_1\), \(\ell_2\), and \(\ell_3\) completely free.

The solid red boundary consists of three segments: (i) when $0\leq r_{AB}\lesssim 0.25$, the boundary is given by \\
$\sqrt{r_{AC}}+2\sqrt{r_{AB}}=1$ and coincides with the left branch of the dashed yellow boundary, (ii) when $0.25\lesssim r_{AB}\lesssim 0.45$, where the boundary is trivial ($r_{AC}=0$), and (iii) when $0.45\lesssim r_{AB}\lesssim 1$, where the boundary coincides with the right branch of the dashed blue boundary. Note that this last boundary 
does not have a closed-form analytical expression, because the constraints on the overlaps lead to transcendental equations similar to Eq. \eqref{eq:transcendental}, where only implicit or numerical solutions are feasible. The experimental results displayed in the left branch of Fig.\ref{fig:regionone} cover segment (i).

\subsection{Violation of $h_n(r)$}
\label{subsec:condition-1b}

In the results of \cite{GalvaoBrod2020, giordani2023experimentalcertification}, the regions which serve as witnesses of coherence and dimension are very distinct, in the sense that the region which is inaccessible to three-qubit states (i.e., close to the point where all three states are orthogonal) is compatible with a set of three classical states in dimension larger than two. Indeed, our experimental witnesses of coherence were obtained in the regime where \(\beta_{ij}=0\), whereas for the dimension witness we used  \(\ell_A = \ell_B = \ell_C\).

We now move to a different set of constraints, studied in \cite{giordani2023experimentalcertification}, namely the \(h_n(r)\) inequalities described in Eq.\ \eqref{hn_coherence_witness}. As discussed previously, these now witness high-dimensional coherence, in some sense, since each \(h_n(r)\) can only be violated by states displaying both basis-independent coherence and a dimension \(n-1\) or greater.

In order to find optimal sets of fractional modes to witness high-dimensional coherence, we numerically maximize the violation of the inequalities defined by the \(h_n(r)\). Assuming the only constraint on the states is a topological charge in \([0,1]\), we find that for \(n=3\) the search recovers exactly the set of states defined in \eqref{caseone}, yielding a violation \(h_3(r)=1.22\), as reported in Table~\ref{tab:overlap_s1}. For \(n=4\), our numerical search identified a set of four fractional states with \(h_4(r)=1.15\) (\(\ell_{1}=0.90\), \(\ell_{2}=\ell_{3}=\ell_{4}=0.60\), \(\beta_{12}=0\), \(\beta_{13}=4.30\,\mathrm{rad}\), \(\beta_{14}=1.97\,\mathrm{rad}\), and \(\beta_{23}=2.33\,\mathrm{rad}\)). Implementing this set experimentally yielded an \(h_4\) violation of approximately 13 standard deviations, as shown in Table~\ref{tab:h4_results}. Note that the set of modes that exhibited a violation of \(h_4\) does not correspond to states with the same symmetry as those considered in either Section~\ref{subsec:condition-2} (with \(\beta_{ij}=0\)) or Section~\ref{subsec:condition-1}. Furthermore, a numerical search found no set of states that would violate \(h_4\) under those symmetry constraints, corroborating the dual role of \(h_{4}\) as a simultaneous dimension and coherence witness.

The next step was to attempt to surpass the bound imposed by $h_5$. However, the numerical search found a maximum value of $h_5=0.85$ (with $\ell_1=\ell_2=\ell_3=\ell_4=0.4$, $\ell_5=0.1$, and $\alpha_1=0\,\mathrm{rad}$, $\alpha_2=1.55\,\mathrm{rad}$, $\alpha_3=3.18\,\mathrm{rad}$, $\alpha_4=4.66\,\mathrm{rad}$, $\alpha_5=4.77\,\mathrm{rad}$). This raises the following question: could a set of five coherent states, with fractional charges in the interval $(0,1)$, be unable to generate a subspace of dimension greater than or equal to $4$? In Appendix~B, we present a purely theoretical argument, based on the Gram matrix and assuming pure states, by which we verify that the modes that maximize $h_5$ exhibit the expected coherence properties and, simultaneously, are linearly independent, spanning a space of dimension $d=5$. Even though this set of states displays the expected coherence and dimensional properties, the absence of a violation of $h_5$ suggests that the family of $h_n$ inequalities is simply not effective at detecting such high-dimensional fractional coherent states, which motivates the search for alternative witnesses compatible with experimental implementation.

\begin{table}[h]
    \centering
    \small
    \caption{Overlap values for the four fractional modes \(\ket{\ell,\beta}\) and the violation of \(h_4\). The experimental values of \(r_{ij}=|\langle \ell _i,\alpha_i|\ell_j,\alpha_j\rangle|^2\) are shown in the second column.}
    \label{tab:h4_results}
    \renewcommand{\arraystretch}{1.1}
    \begin{tabular}{@{}l c@{} c@{}}
        \hline\hline
                    & \hspace{0.5cm}$r_{the}$ &\hspace{0.5cm} $r_{exp}$ \\
        \(r_{01}\) & \hspace{0.5cm}\(0.72\) &\hspace{0.5cm} \(0.67\pm0.01\) \\
        \(r_{02}\) & \hspace{0.5cm}\(0.50\) & \hspace{0.5cm}\(0.52\pm0.01\) \\
        \(r_{03}\) & \hspace{0.5cm}\(0.50\) & \hspace{0.5cm}\(0.50\pm0.01\) \\
        \(r_{12}\) & \hspace{0.5cm}\(0.16\) & \hspace{0.5cm}\(0.16\pm0.01\) \\
        \(r_{13}\) & \hspace{0.5cm}\(0.21\) & \hspace{0.5cm}\(0.17\pm0.01\) \\
        \(r_{23}\) & \hspace{0.5cm}\(0.17\) & \hspace{0.5cm}\(0.23\pm0.01\) \\
        \(h_4\)  & \hspace{0.5cm}\(1.15\) &\hspace{0.5cm} \(1.13\pm0.01\) \\
        \hline\hline
    \end{tabular}
\end{table}

\section{Conclusion}
\label{sec:conclusion}

In this work, we analyze families of light modes carrying fractional orbital angular momentum (OAM) by means of the basis-independent witnesses of coherence and dimensionality based on overlap measurements, and introduced in \cite{GalvaoBrod2020, giordani2023experimentalcertification}. Such witnesses are appealing because they allow experimental certification of the modes’ properties without assuming state purity, besides being robust in the sense of remaining effective even under small deviations from the theoretically expected behavior. To perform the measurements, we employ an interferometric technique based on the transverse superposition of their intensity distributions. The resulting interference pattern encodes all the information required so that the mode overlap between pairs of spatial modes can be obtained directly through a simple two-dimensional Fourier transform.  
The method is extremely robust to relative phase fluctuations, eliminating the phase-locking requirement typical of standard interferometers. In this way, it becomes feasible to investigate the relational properties of spatial modes, which can be obtained and analyzed through overlap measurements between them, regardless of the modes’ complexity.

We have observed that fractional OAM modes present a diversity of behaviors. By a careful choice of the parameters, we showed experimentally that they can display basis-independent coherence and high dimensionality even when restricted to $0 \leq \ell \leq 1$. Interestingly, as we investigated numerically, this range supports neither a triplet of mutually orthogonal states, nor a set of three linearly-dependent states within it. Moreover, the witnesses considered were not able to experimentally certify a set of fractional modes with coherence and dimension $\geq 4$, although theoretical arguments point to their existence. Thus, the search for new criteria that are experimentally sensitive to such high-dimensional coherent modes remains open.

Beyond the fundamental interest of situating fractional vortices within the landscape of quantum resources, our results on the relational properties between fractional modes, namely effective dimension and coherence, pave the way for applications in practical, high-dimensional quantum-information protocols \cite{forbes2021structured,mafu2013higher,sun2024experimental, rubinsztein2016roadmap}. As a direction for future research, we believe it would be fruitful to develop a similar interferometric approach to measure higher-order Bargmann invariants, which capture relational properties of more than two states at a time  \cite{oszmaniec2024measuring}. This would provide more complete information on the structure of fractional OAM modes and their capabilities for quantum information processing tasks. This seems like a promising direction, given some recent work characterizing out-of-time-order correlators (OTOCs) \cite{wagner2024quantum}, weak values \cite{wagner2023simple} and quasi-probability distributions \cite{schmid2024kirkwooddirac} using such higher-order unitary invariants.

\section*{Acknowledgments}
	
 Funding was provided by Conselho Nacional de Desenvolvimento Científico e Tecnológico (CNPq), 
Coordenação de Aperfeiçoamento de Pessoal de Nível Superior (CAPES), Fundação Carlos Chagas 
 Filho de Amparo à Pesquisa do Estado do Rio de Janeiro (FAPERJ), Instituto Nacional de Ciência e Tecnologia de Informação Quântica (INCT-IQ 465469/2014-0), Fundação de Amparo à Pesquisa do Estado de São Paulo (FAPESP, grants 2021/06823-5, 2022/15036-0 and 2022/15035-3), and the ERC Advanced Grant QU-BOSS, GA no. 884676. E. F. G.  acknowledges support from FCT – Fundaç\~{a}o para a Ciência e a Tecnologia (Portugal) via project CEECINST/00062/2018. 
	

\begin{widetext}	
\section*{Appendix A: Derivation of the overlap expression}
\label{derivation_of_the_overlap}

		Let us assume, without loss of generality, that $\alpha\geq\alpha'$. The complementary case can be easily inferred by complex conjugation of the overlap expression. In this case, the overlap reads
	\begin{eqnarray}
    \label{eq:inner_product}
		\braket{\ell',\alpha'}{\ell,\alpha} &=& 
		\frac{1}{2\pi} \int_0^{\alpha'} f^*_{\ell',\alpha'}(\phi)\, f_{\ell,\alpha}(\phi) \,d\phi +
		\frac{1}{2\pi} \int_{\alpha'}^{\alpha} f^*_{\ell',\alpha'}(\phi)\, f_{\ell,\alpha}(\phi) \,d\phi +
		\frac{1}{2\pi} \int_{\alpha}^{2\pi} f^*_{\ell',\alpha'}(\phi)\, f_{\ell,\alpha}(\phi) \,d\phi
		\nonumber\\
		&=& \frac{e^{-i(\ell\alpha - \ell'\alpha')}}{2\pi i\,(\ell-\ell')}
		\left[
		e^{i2\pi(\ell-\ell')}\,\left( e^{i(\ell-\ell')\alpha'}-1\right) +
		e^{-i2\pi\ell'}\,\left(
		e^{i(\ell-\ell')\alpha}-e^{i(\ell-\ell')\alpha'}\right) + 
		e^{i2\pi(\ell-\ell')}-e^{i(\ell-\ell')\alpha}\right]
		\nonumber\\
		&=& \frac{e^{-i\ell'(\beta + \pi)}}{\pi(\ell-\ell')}
		\left[
		e^{i(\pi-\beta)(\ell-\ell')}\sin(\ell\pi)-\sin(\ell'\pi)\right]
        \nonumber\\
		&=& \frac{e^{-2i\ell' \pi}}{\pi(\ell-\ell')}
		\left[
		e^{i(\pi-\beta)\ell}\sin(\ell\pi) - e^{i(\pi-\beta)\ell'}\sin(\ell'\pi)\right],
		\label{eq:fracOverlap3}
	\end{eqnarray}
	where $\beta = \alpha - \alpha'\,$. The squared absolute value of the overlap is readily obtained as
	\begin{eqnarray}
		\abs{\braket{\ell',\alpha'}{\ell,\alpha}}^2 = 
		\frac{1}{\pi^2(\ell-\ell')^2}\, 
		\left[ \sin^2(\pi\ell) + \sin^2(\pi\ell') 
		-2 \cos[(\pi-\beta)(\ell-\ell')]\sin(\pi\ell)\,\sin(\pi\ell')\right]\,.
		\label{eq:AbsFracOverlap}
	\end{eqnarray}
	Note that the following property holds: $\abs{\braket{\ell',\alpha'}{\ell,\alpha}}^2 = 
	\abs{\braket{\ell',0}{\ell,\beta}}^2$.
	
	The following trigonometric relations, 
	\begin{eqnarray}
		&&	\!\!\!\!\! \sin^2 x + \sin^2 y = \sin^2 (x-y) + 2\cos(x-y)\sin x \sin y\,,
		\nonumber\\
		&&	\!\!\!\!\! \cos(x\!-\!y) - \cos(x\!-\!y\!-\!\theta) = -2\sin(\theta/2) \sin(x\!-\!y\!-\!\theta/2)\,,
		\nonumber
	\end{eqnarray}
	with $x = \ell\pi\,$, $y = \ell'\pi\,$, and $\theta = (\ell - \ell')\beta\,$, can be used to put the expressions 
	for the overlaps in the form of cardinal sine functions of $\ell-\ell'$, allowing for easy calculation of the limit 
	$\ell\to\ell'\,$. This brings us to 
	\begin{equation}
		\abs{\braket{\ell',\alpha'}{\ell,\alpha}}^2 = 
		\mathrm{sinc}^2[(\ell - \ell')\pi] - \frac{\beta(2\pi-\beta)}{\pi^2} \sin(\ell\pi)\sin(\ell'\pi)
		\mathrm{sinc}{\left[(\ell - \ell')\frac{\beta}{2}\right]}
		\mathrm{sinc}\left[(\ell - \ell')\left(\pi-\frac{\beta}{2}\right)\right]\;.
				\label{eq:AbsFracOverlapSinc}
	\end{equation}
	Equations \eqref{eq:AbsFracOverlap} and \eqref{eq:AbsFracOverlapSinc} will be used to compute the overlap between 
	different fractional OAM states.
\end{widetext}

\begin{widetext}	
\section*{Appendix B: Evaluation of Linear Independence Using the Gram Matrix}
\label{Gram_matrix}

We now present a theoretical argument to certify that the modes obtained by optimizing the function $h_5$ ($\ell_0=\ell_1=\ell_2=\ell_3=0.4$, $\ell_4=0.1$, and
$\alpha_0=0\,\mathrm{rad}$, $\alpha_1=1.55\,\mathrm{rad}$, $\alpha_2=3.18\,\mathrm{rad}$, $\alpha_3=4.66\,\mathrm{rad}$, $\alpha_4=4.77\,\mathrm{rad}$),
which we denote by the set $\{\ket{\ell_i,\alpha_i}\}_{h_5}$, are indeed coherent and span a 5-dimensional space, despite not being detected by the $h_5$ criterion.

The assessment of linear independence can be carried out by means of the Gram matrix, presented below. Consider a set of pure states $\{\ket{\psi_i}\}_{i=0}^{n}$, associated with it is the Gram matrix, defined by

\begin{equation}
\label{eq:gram-def}
G_{ij}=\langle \psi_i \mid \psi_j \rangle,\qquad i,j=0,\dots,n\,,
\end{equation}

which is Hermitian and positive semidefinite; for normalized states, $G_{ii}=1$. As shown in~\cite{oszmaniec2024measuring}, the states are linearly independent if and only if the determinant of the Gram matrix is strictly positive ($\det G>0$). We emphasize that this is a purely theoretical assessment, since it assumes pure states and the entries of $G$ involve inner products that are, in general, not directly observable. Nevertheless, when analytical expressions for these inner products are available, as in the case of fractional modes (Eq. \eqref{eq:inner_product}), the Gram matrix is a useful tool for inferring the coherence and linear independence of the set.

Thus, for the set of fractional modes $\{\ket{\ell_i,\alpha_i}\}_{h_5}$, the determinant of its Gram matrix is readily computed, yielding $\det G=0.005$, which guarantees that the five modes are not linearly dependent and therefore span a 5-dimensional space. Moreover, in this idealized setting, coherence between the states is certified by the fact that all pairwise overlaps satisfy $0<r_{ij}<1$ for every $i\neq j$, as indicated in Table~\ref{tab:modes_h_5}. Hence, we can assert that these modes are simultaneously coherent and span a 5-dimensional space, although they are not detected by the experimentally robust $h_5$ witness.

\begin{table}[h]
    \centering
    \small
    \caption{The set of modes represented by $\{\ket{\ell_i,\alpha_i}\}_{h_5}$ has parameters
$\ell_0=\ell_1=\ell_2=\ell_3=0.4$, $\ell_4=0.1$, $\alpha_0=0\,\mathrm{rad}$, $\alpha_1=1.55\,\mathrm{rad}$, $\alpha_2=3.18\,\mathrm{rad}$, $\alpha_3=4.66\,\mathrm{rad}$, and $\alpha_4=4.77\,\mathrm{rad}$. These values were obtained via a numerical search aimed at violating the function $h_5$. Below, we present the angular differences $\beta_{ij}$ for each pair of fractional modes, together with their respective theoretical overlap values $r_{ij}$.
}
    \label{tab:modes_h_5}
    \renewcommand{\arraystretch}{1.1}
    \begin{tabular}{@{}l c@{\hspace{0.7cm}} c@{}}
        \hline\hline
         & $\beta_{ij}$ [rad] & $r_{the}$ \\
        \hline
        $r_{01}$ & 1.55 & 0.328 \\
        $r_{02}$ & 3.18 & 0.096 \\
        $r_{03}$ & 4.66 & 0.307 \\
        $r_{04}$ & 4.77 & 0.541 \\
        $r_{12}$ & 1.63 & 0.305 \\
        $r_{13}$ & 3.11 & 0.096 \\
        $r_{14}$ & 3.22 & 0.464 \\
        $r_{23}$ & 1.48 & 0.349 \\
        $r_{24}$ & 1.59 & 0.534 \\
        $r_{34}$ & 0.11 & 0.719 \\
        \hline\hline
    \end{tabular}
\end{table}

 \end{widetext}   
\bibliography{frac-oam-bibifile}	
\end{document}